\documentclass[
reprint, 
superscriptaddress,
 amsmath,amssymb,
 aps,
]{revtex4-2}

\usepackage[utf8]{inputenc}
\usepackage{graphicx}
\usepackage{dcolumn}
\usepackage{bm}
\usepackage{hyperref}
\usepackage[mathlines]{lineno}
\usepackage{amsmath} 
\usepackage{siunitx}   
\usepackage{multirow}  
\usepackage{physics}
\usepackage{xcolor}
\usepackage{comment} 
\usepackage{physics}
\usepackage{float}

\begin{document}

\title{Highly efficient microwave storage and retrieval using a superconducting chiral $\Lambda$-type molecule}

\author{Kai-I Chu}
\email[Correspondence to: ]{kaiichu0903@gmail.com}
\affiliation{Department of Physics, National Central University, Taoyuan City 32001, Taiwan}

\author{Yung-Fu~Chen}
\affiliation{Department of Physics, National Central University, Taoyuan City 32001, Taiwan}
\affiliation{Quantum Technology Center, National Central University, Taoyuan City 32001, Taiwan}

\author{Wen-Te Liao}
\email[Correspondence to: ]{wente.liao@g.ncu.edu.tw}
\affiliation{Department of Physics, National Central University, Taoyuan City 32001, Taiwan}
\affiliation{Quantum Technology Center, National Central University, Taoyuan City 32001, Taiwan}
\affiliation{Physics Division, National Center for Theoretical Sciences, Taipei 10617, Taiwan}

\begin{abstract}

We theoretically investigate a high-efficiency and broadband microwave storage and retrieval scheme employing a superconducting artificial chiral molecule embedded in a one-dimensional transmission line. By optimizing the parametric coupling, the chiral $\Lambda$-type molecule enables near 100\% storage efficiency and fidelity across a bandwidth of 100 MHz. Our results provide a feasible pathway toward implementing microwave quantum memories compatible with broadband quantum networks.

\end{abstract}

\date{\today}

\maketitle

The superconducting circuit platform is a leading approach for realizing noisy intermediate-scale quantum processors \cite{devoret2013superconducting,arute2019quantum,kjaergaard2020superconducting,gao2025establishing}. To achieve fault-tolerant and scalable quantum computation, it is crucial to develop a quantum network \cite{kimble2008quantum,wehner2018quantum,wei2022towards} that interconnects multiple moderate-scale quantum processors through quantum communication and entanglement, enhancing computational capabilities and improving hardware efficiency. Within such a superconducting circuit-based network \cite{magnard2020microwave}, microwave quantum memory, capable of storing and retrieving propagating quantum information, will be a key building block \cite{lvovsky2009optical,heshami2016quantum,liu2023quantum}. It enables synchronization and buffering of quantum information for parallel processing, improves computational efficiency, and facilitates quantum state transfer over intermediate distances in distributed superconducting quantum systems \cite{caleffi2024distributed}. Additionally, it can support long-distance quantum teleportation \cite{duan2001long} when required for large-scale quantum networks.

Recent approaches to microwave memory based on superconducting circuits coupling to high-coherence resonators can be categorized into two main schemes. The first approach invokes the atomic frequency comb, where an array of resonators absorbs the probe pulse \cite{bao2021demand,matanin2023toward,makihara2024parametrically}. Storage is achieved by either turning off the coupling between the information channel and the resonators \cite{makihara2024parametrically} or by aligning the resonator frequencies \cite{bao2021demand}. The stored pulse is retrieved through an echo rephasing process. An efficiency of up to 60\% was realized in Ref.~\cite{matanin2023toward} using a fixed-delay design, while more recent work has enabled on-demand storage and retrieval by incorporating an additional switch~\cite{matanin2026superconducting,perminov2023integrated,perminov2024multiresonator}. However, even if the resonators exhibit high coherence, the storage efficiency and fidelity are limited by deviations in the coupling strengths between the resonators and the information channel, variations in resonator frequencies and intrinsic losses, and impedance-matching conditions.  Another challenge is that these memory devices typically require an additional circulator to transfer the stored information to another channel, which causes additional losses and further degrades the stored information \cite{moiseev2018broadband}. The second approach relies on electromagnetically induced transparency (EIT) using a three-level $\Lambda$-type superconducting-qubit-resonator system \cite{chu2025slow}. The EIT mechanism slows down a probe pulse and renders the system transparent under the influence of the coupling field, due to quantum interference \cite{fleischhauer2000dark,phillips2001storage,fleischhauer2005electromagnetically}. The probe pulse is stored in and retrieved from the high-coherence resonator by adiabatically switching the coupling field off and on. However, this single atom, embedded in an open transmission line, has an effective optical depth of only 2, and its theoretical maximum storage and retrieval efficiency is further reduced to 25\% due to bidirectional coupling \cite{chu2023three}. Note that highly efficient EIT quantum memory generally requires a large optical depth \cite{hsiao2018highly,gorshkov2007photon}. While increasing the number of artificial atoms can enhance optical depth, it also introduces cooperative decoherence, making the approach challenging.

In this work, we propose the first $\Lambda$-type microwave memory scheme using a superconducting artificial chiral molecule embedded in a one-dimensional open transmission line to address the above challenges. Here, chirality refers to the fact that the atomic scattering of the probe field depends on its propagation direction \cite{lodahl2017chiral,joshi2023resonance}. Our optimal control protocol broadens the frequency bandwidth in contrast to the original giant unidirectional emitter proposals \cite{gheeraert2020programmable,guimond2020unidirectional,kannan2023demand}, where directional absorption and emission are realized, but the operating bandwidth is inherently constrained by phase matching. Under the two-photon resonant condition, the molecule with the weak probe and the coupling fields forms the dark state (see Appendix \ref{appendixA4}), and the probe is adiabatically transferred to the memory qubit state by controlling the coupling field, enabling microwave storage. The protocol operates in both the resonant EIT regime and the detuned Raman regime. In the EIT regime, high storage efficiencies $> 99\%$ can be achieved without requiring a large number of atoms. In the Raman regime, comparable efficiencies are maintained with storage bandwidths of around 100 MHz, thereby overcoming the bandwidth limitation imposed by phase-matching, albeit at the cost of requiring a stronger coupling Rabi frequency. The scheme exhibits a large tolerance to emission-frequency mismatches between quantum nodes, which is useful for quantum networks. To our knowledge, this represents the first proposal for realizing Raman-style storage in a fully superconducting system, while retaining the directional control advantages inherent to chiral architectures \cite{gheeraert2020programmable, guimond2020unidirectional, kannan2023demand}. Both operations can achieve near-unity fidelity across a broad range of pulse durations. In contrast to atomic ensembles \cite{hsiao2018highly,gorshkov2007photon}, our single artificial-molecule scheme achieves high storage efficiency independent of optical depth. Our theory provides a general framework for any chiral $\Lambda$-type system, given in Appendix \ref{appendixA4}.

We also present a simple adiabatic control, suitable for the EIT regime, that still achieves high-efficiency microwave storage and retrieval. Moreover, our scheme does not need any circulator since the atom is embedded in an open transmission line. The propagating direction of the retrieved pulse can be controlled by adjusting the phase of the coupling field, which offers the versatility of our scheme for quantum memory applications. 

\begin{figure}[t!]
    \centering
    \includegraphics[width=0.4\textwidth]{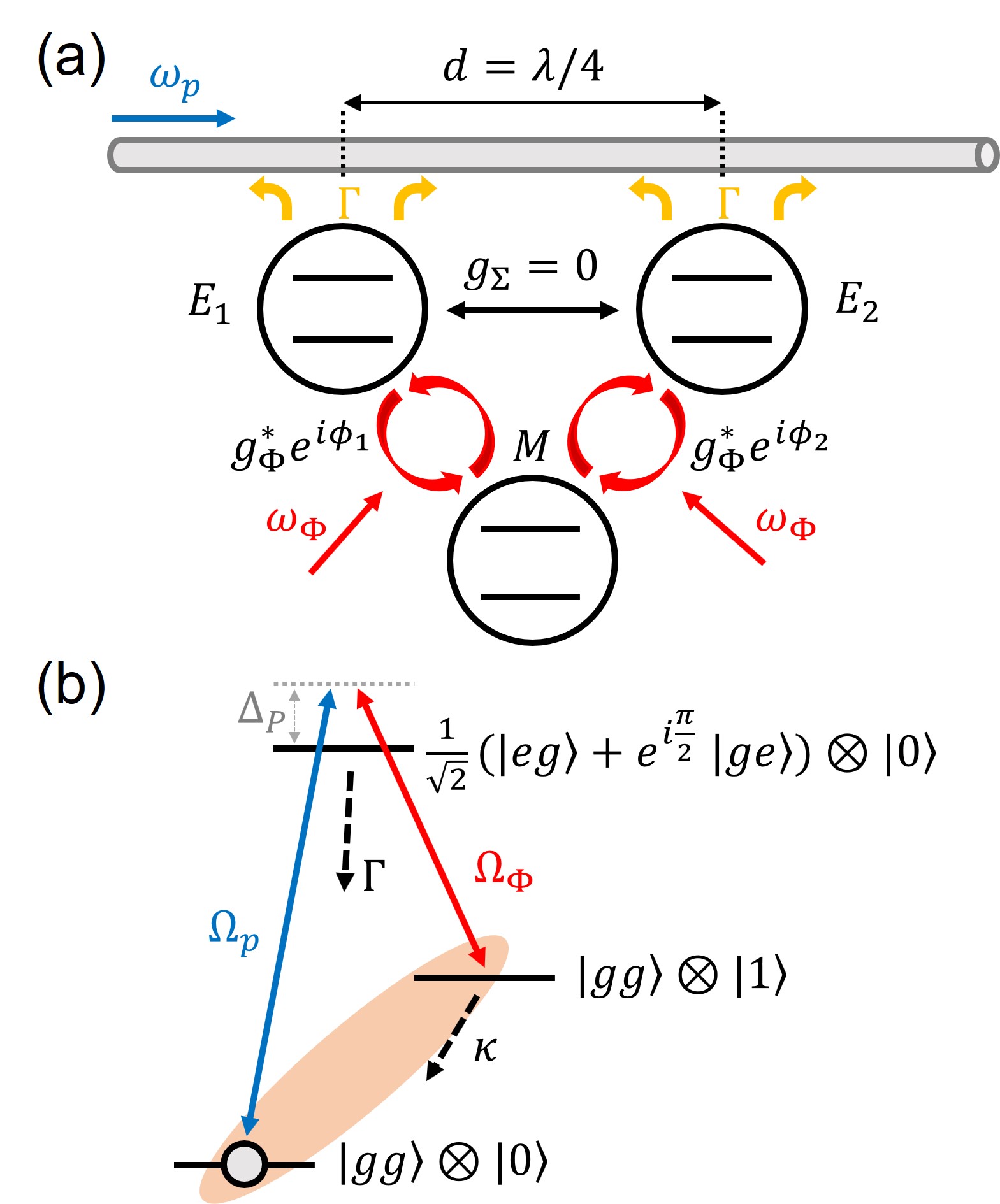}
    \caption{ (a) Schematic diagram of an artificial chiral molecule embedded in a one-dimensional transmission line. 
    The artificial molecule consists of a pair of resonant emitter qubits $E_1$ and $E_2$ and a detuned memory qubit $M$.
    $E_1$ and $E_2$ are separated by $\lambda/4$ with zero interaction ($g_{\Sigma} = 0$), and they couple to $M$ with controllable effective coupling strength $g_{\Phi}^{*}e^{i\phi_{1(2)}}$ through the parametric modulations. (b) Energy-level diagram of the lowest three states of the chiral molecule, forming an effective $\Lambda$-type system. The ground, excited, and metastable states are denoted as $\ket{gg}\otimes\ket{0}$, $(\ket{eg} + e^{i\pi/2} \ket{ge})/\sqrt{2} \otimes \ket{0}$, and $\ket{gg}\otimes\ket{1}$, respectively. A probe field is injected into the transmission line and couples to both emitters with Rabi frequency $\Omega_{p}$ and detuning $\Delta_p$. The parametric modulations generate a coupling Rabi frequency $\Omega_{\Phi}$ with the same detuning $\Delta_p$, enabling the transition between the excited and metastable states. The $\Lambda$-type configuration driven by the probe and coupling fields supports the formation of a dark-state coherence (orange).}
    \label{fig:setup}
\end{figure}

Our scheme is illustrated in Fig. \ref{fig:setup}(a). A pair of resonant emitter qubits, $E_1$ and $E_2$, with transition frequency $\omega_{e}$, couples to an open transmission line with a separation distance $d=\lambda/4$, where the wavelength $\lambda =2\pi c/\omega_{e}$. This quarter-wavelength spacing prevents cooperative effects while maximizing the exchange coupling to $J = \Gamma/2$ \cite{lalumiere2013input}, where $\Gamma$ is the emitter relaxation rate. We assume that this exchange coupling $J$ can be fully canceled by introducing a tunable coupler \cite{yan2018tunable,li2020tunable}, leading to a total coupling of $g_{\Sigma} = 0$. The absence of coupling results in fully directional emission and absorption of the emitter pair \cite{gheeraert2020programmable,guimond2020unidirectional,kannan2023demand}. In the weak probe regime, the excited state forms an entangled state, $(\ket{eg} + e^{i\pi/2} \ket{ge})/\sqrt{2}$, where $\ket{g}$ and $\ket{e}$ denote the ground and excited states of each emitter qubit, respectively, due to time-reversal symmetry \cite{gheeraert2020programmable}. The phase sign is determined by the propagation direction of the probe field. Here, the probe is incident from the left. Additionally, each emitter is coupled to the same detuned high-coherence memory qubit or resonator, $M$, with transition frequency $\omega_{m}$, positioned far from the transmission line. The coupling strength between the emitters and the memory qubit is smaller than their frequency difference, minimizing energy transfer from the memory qubit through the emitters to the transmission line. We treat these three quantum objects as a single chiral molecule. 
The lowest three energy levels of the molecule, depicted in Fig. \ref{fig:setup}(b), form an effective $\Lambda$-type structure with states $\ket{gg}\otimes\ket{0}$, $(\ket{eg} + e^{i\pi/2} \ket{ge})/\sqrt{2} \otimes \ket{0}$, and $\ket{gg}\otimes\ket{1}$, where $\ket{0}$ and $\ket{1}$ denote the ground and excited states of the memory qubit, respectively. The state $\ket{gg}\otimes\ket{1}$ serves as the metastable state of the molecule, which is used for microwave storage.

To realize a $\Lambda$-type configuration that enables mapping of the weak probe field with one-photon detuning $\Delta_p=\omega_e-\omega_p$ and Rabi frequency $\Omega_p$ onto the coherence of the metastable state, the otherwise forbidden transition between the emitters and the memory qubit must be activated \cite{blais2007quantum}. This is achieved using a parametric modulation technique \cite{strand2013first,mckay2016universal,caldwell2018parametrically}, which modulates the emitters' transition frequency with frequency $\omega_{\Phi}$, such that the one-photon detuning $\Delta_\Phi \equiv \omega_e - \omega_m - \omega_\Phi$ matches $\Delta_p$.
For the resonant modulation, it can induce an effective coupling strength $g_{\Phi}^*e^{i\phi_{1(2)}}$ between the emitters and the memory qubit, with $\phi_{1(2)}$ the respective modulation phases. The derivation of the parametric modulation effect is presented in Appendix \ref{appendixA1}. Furthermore, the phase difference must be set to $\phi \equiv \phi_{1} - \phi_{2} = -\pi/2$, ensuring a non-reciprocal effect \cite{cao2024parametrically} for the resonant probe incident from the left. The emission from the emitters occurs only to the right due to the interference effect governed by $\phi$ and the propagation phase accumulation over $d=\lambda/4$. For a large one-photon detuning (Raman regime), the same phase relation remains effective, with only a small residual loss that is negligible for storage bandwidths on the order of 100 MHz, as demonstrated later. As a result, these two modulations collectively act as a coupling field, resulting in an effective Rabi frequency $\Omega_{\Phi}=2\sqrt{2}|g_{\Phi}|$, which facilitates the transition from the excited state to the metastable state of the chiral molecule (see Appendix \ref{AppendixA3}).

\begin{figure}[t!]
    \centering
    \includegraphics[width=0.4\textwidth]{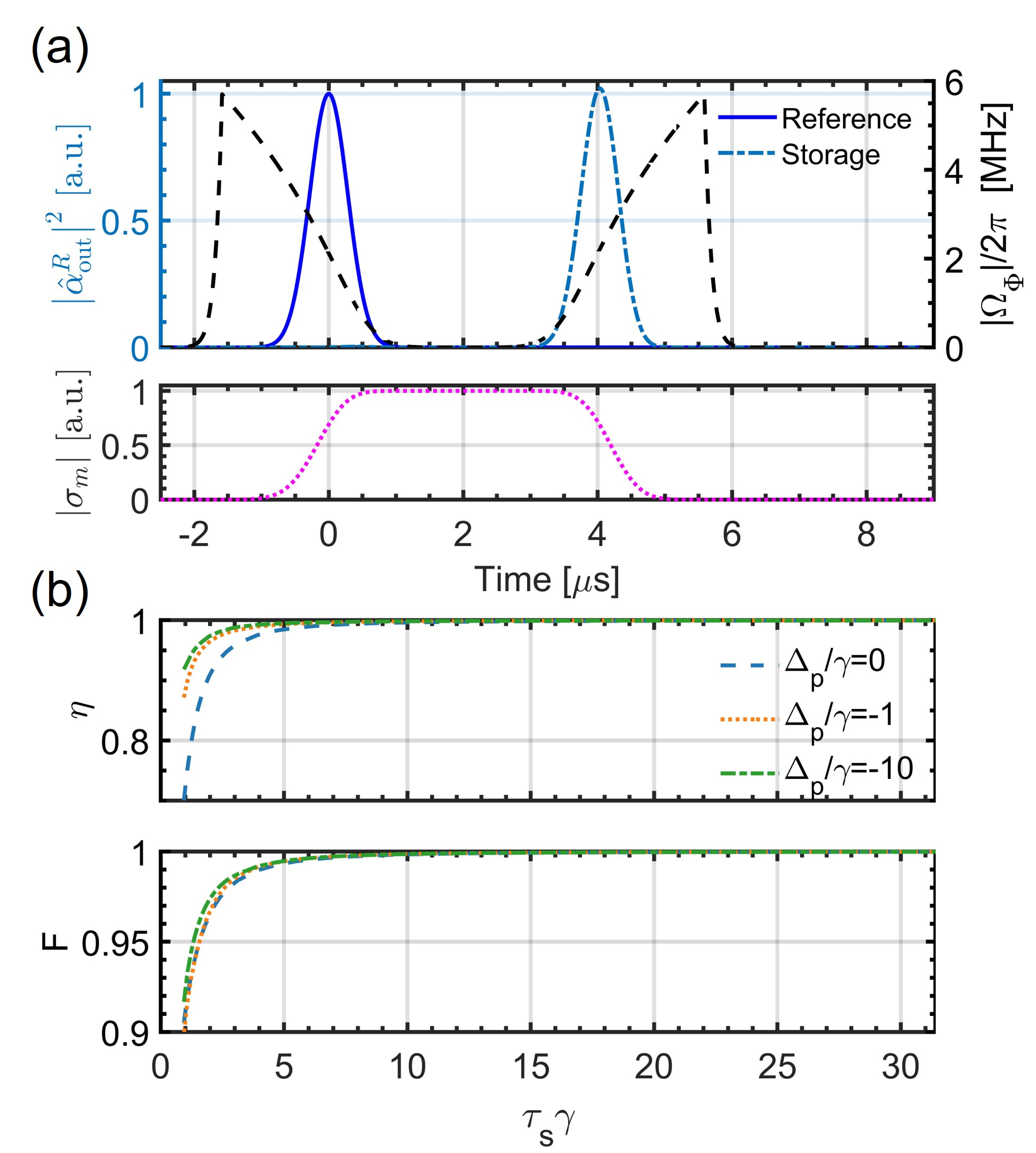}
    \caption{(a) Optimized storage and retrieval sequence. The upper plot shows the output probe signal $|\alpha_{\text{out}}|^2$. The black-dashed line represents the control sequence of the coupling field. The blue line is the input probe pulse. The light-blue dot-dashed line depicts the retrieved pulse. The lower plot shows the transformation between the microwave coherence and the memory qubit coherence $\sigma_m$. (b) Storage and retrieval efficiency $\eta$ and fidelity $\text{F}$ as functions of pulse duration $\tau_s$ and $\Delta_p$. Different colors correspond to different values of $\Delta_p$.}
    \label{fig:2}
\end{figure}

The Hamiltonian describing the interaction between the probe field and the chiral molecule under the coupling field, within the rotating frame approximation \cite{kannan2023demand,kervinen2019landau}, is explained in Appendix \ref{appendixA1}--\ref{AppendixA3} and given by
\begin{equation}
    \begin{aligned}
    \label{eq:Ham eq}
    H = &\sum_{i=1,2} \Delta_p \sigma_{i}^\dagger \sigma_{i} + \delta \sigma_{m}^\dagger \sigma_{m} \\ &+ g_{\Phi}^*(e^{i\phi} \sigma_{1}^\dagger \sigma_{m} + \sigma_{2}^\dagger \sigma_{m}) + \frac{\Omega_p}{2} (\sigma_{1}^\dagger + e^{ikd} \sigma_{2}^\dagger) + \text{H.c.}, 
    \end{aligned}
\end{equation}
where $\sigma_{1(2)}$ denotes the lowering operator for the emitter 1(2), $\sigma_{m}$ is the lowering operator for the memory qubit, $\delta =\Delta_p-\Delta_{\Phi}$ is the two-photon detuning, and $k$ is the wavevector of the probe field. The Markovian approximation is applied and valid when $\Gamma \tau_p \ll 1$~\cite{frisk2014designing,guo2017giant,frisk2020quantum,sinha2020non,lalumiere2013input}, where $\tau_p$ is the propagation time between the two emitter positions. This can be understood from the fact that the emitter operators change negligibly during the propagation time and can therefore be approximated as undergoing free evolution~\cite{lalumiere2013input}. When the pulse duration $\tau_s$ is larger than $\tau_p$~\cite{almanakly2025deterministic} and the long-time approximation $\omega_e\tau_s\gg1$~\cite{lalumiere2013input}, the two emitters experience the same field amplitude but acquire different phases~\cite{gheeraert2020programmable,guimond2020unidirectional}. This approach has been widely used and experimentally verified in many works, e.g., Ref.~\cite{joshi2023resonance,cao2024parametrically,almanakly2025deterministic}. In a typical waveguide-QED setting, taking $d \approx 1 \mathrm{cm}$, a microwave velocity $v \approx 2 \times 10^8 \mathrm{m/s}$, and $\Gamma/2\pi \approx 10 \mathrm{MHz}$, one obtains $\Gamma \tau_p \approx 10^{-3}$, which satisfies this condition well. When this condition is not satisfied, the system enters a non-Markovian regime, where the finite propagation time gives rise to delayed feedback of the emitted field. One may also consider a more general situation in which the emitters experience different field amplitudes, as discussed in Ref.~\cite{perminov2019spectrally}. 

The time evolution of the system is governed by the Lindblad master equation,
\begin{equation}
\begin{aligned}
    \label{eq:Master eq}
    \dot{\rho}=&-i[H,\rho] + \sum_{i=1,2} \left( \frac{\Gamma}{2} \mathcal{D}[\sigma_i]\rho
+ \gamma_{\phi} \mathcal{D}[\sigma_{i}^\dagger\sigma_{i}]\rho \right) + \frac{\Gamma_{m}}{2} \mathcal{D}[\sigma_m]\rho,
\end{aligned}
\end{equation}
where $\mathcal{D}[\mathcal{O}]\rho=2\mathcal{O}\rho\mathcal{O}^\dagger-\rho\mathcal{O}^\dagger\mathcal{O}-\mathcal{O}^\dagger\mathcal{O}\rho$  is the dissipator, and $\rho(t)$ is the system’s density matrix. $\gamma_{\phi}$ and $\Gamma_{m}$ are the emitters' pure dephasing rate and the memory qubit's loss rate, respectively. The dynamics of the density matrix is solved using QuTiP \cite{johansson2012qutip}. To determine the output field propagating to the right, we use the input-output formalism, where the output coherent field is given by \cite{lalumiere2013input,gheeraert2020programmable,kannan2023demand}
\begin{equation}
\begin{aligned}
    \label{eq:rfiled}
\hat{\alpha}^R_{out}= \hat{\alpha}^R_{in} - i\sqrt{\frac{\Gamma}{2}} ( \sigma_1 + \sigma_2 e^{-ikd} ). 
\end{aligned}
\end{equation}
The input coherent field is given by $\hat{\alpha}^R_{in}=\Omega_p/\sqrt{2\Gamma}$. The steady-state transmission coefficient and the slow light results are discussed in Appendix \ref{AppendixB}.

\begin{figure}[t!]
    \centering
    \includegraphics[width=0.4\textwidth]{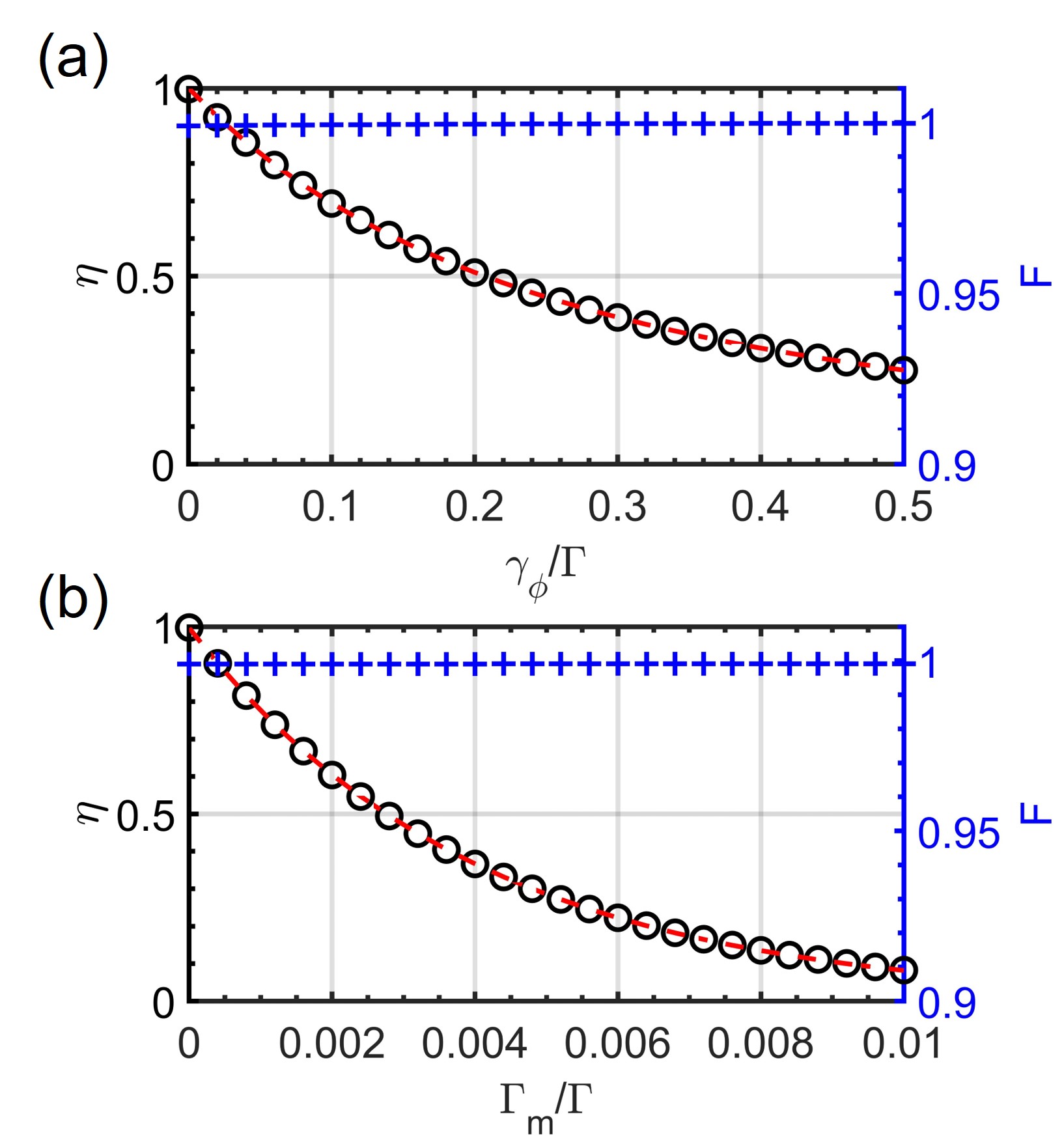}
    \caption{(a) $\eta$ (black open circles) and $\text{F}$ (blue crosses) as functions of pure dephasing rate $\gamma_{\phi}$. The red dashed line shows the analytical curve.
    (b) $\eta$ and $\text{F}$ as functions of memory decay rate $\Gamma_m$. The red dashed line denotes the fitting curve.}  
    \label{fig:3}
\end{figure}

For any ideal chiral system without the phase-matching condition, we demonstrate that the storage and retrieval efficiency $\eta$ can attain 100\%, remaining independent of the optical depth, $\Delta_p$, and $\Omega_{\Phi}$. The derivation is provided in Appendix \ref{AppendixC}. By defining the chiral operator $\sigma_R=(\sigma_1-i\sigma_2)/\sqrt{2}$ and assuming that the time evolution satisfies both the adiabatic condition $\dot{\sigma_R}\approx0$ and the temporal retrieval-matching condition $\hat{\alpha}^R_{out}(t)=\sqrt{\eta}\hat{\alpha}^R_{in}(t-t_r)$, we obtain the optimal control function for retrieving the stored probe field (see Appendix \ref{AppendixD}), expressed as
\begin{equation}
\begin{aligned}
    \label{eq:opt_main}
g_{\Phi}(t)=-\frac{1}{2}\frac{i\Delta_p+\gamma}{\sqrt{\gamma}}\frac{\hat{\alpha}^R_{in}(t-t_r)}{\sqrt{S_{r}}} e^{i\frac{\Delta_p}{2\gamma}\ln{S_{r}}},
\end{aligned}
\end{equation}
where $S_{r}=\int_{t}^{\infty} |\hat{\alpha}^R_{in}(\tau-t_r)|^2 d\tau$ with $t_r$ as the beginning of the retrieval process and $\gamma=\gamma_{\phi}+\Gamma/2$. The optimal control function for storage is simply the time reversal of Eq. \ref{eq:opt_main}, with $S_{r}$ replaced by $S_{s}=\int_{-\infty}^{t} |\hat{\alpha}^R_{in}(\tau)|^2 d\tau$. The adiabatic condition further requires that $\tau_s\gamma\gg1$, as derived in Appendix \ref{AppendixD1}.

For the numerical simulations, the system parameters were chosen as $\omega_e/2\pi = 5$ GHz, $\omega_m/2\pi = 4$ GHz, and $\Gamma/2\pi = 10$ MHz, which are typical in waveguide quantum electrodynamics. The optimal resonant light storage and retrieval process of a Gaussian probe pulse, $\Omega_p(t)=\Omega_p\exp(-t^2/2{\tau_s}^2)$, with $\Omega_p=0.001\Gamma$ and duration of $\tau_s=400$ ns, is illustrated in Fig. \ref{fig:2}(a). As the coupling field is adiabatically turned off, the probe pulse is mapped onto the coherence of the metastable state, demonstrating the dark state transfer mechanism. Fig. \ref{fig:2}(b) presents the efficiency $\eta$, defined as
\begin{equation}
    \label{eq:efficiency}
    \eta = \frac{\int_{t_{r}}^{\infty} |\hat{\alpha}^R_{out}|^2 dt}{\int_{-\infty}^{\infty} |\hat{\alpha}^R_{in}|^2 dt},
\end{equation}
as a function of $\tau_s$ and $\Delta_p$ under ideal conditions (no dephasing or loss). Nearly 100\% efficiency is observed regardless of $\Delta_p$, confirming that the probe pulse is fully compressed into the atom provided that $\tau_s \gamma \gg 1$. The classical fidelity $\text{F}$, defined as the similarity between the retrieved and input pulses, is given by \cite{chen2013coherent}  
\begin{equation}
    \label{eq:efficiency}
    \text{F} = \frac{|\int \hat{\alpha}^{R*}_{in}(t)\hat{\alpha}^R_{out}(t-t') dt|^2}{\int|\hat{\alpha}^R_{in}|^2 dt\int|\hat{\alpha}^R_{out}|^2 dt},
\end{equation}
where $t'$ denotes the time of the maximum amplitude of the retrieved pulse. As shown in Fig. \ref{fig:2}(b), the fidelity also approaches unity. Importantly, in the Raman regime, the magnitude of $g_{\Phi}$ scales with $\Delta_p$. While the storage bandwidth can still be increased, $g_{\Phi}$ is ultimately constrained by the bare coupling strength between the emitters and the memory qubit (see Appendix \ref{appendixA1}). Moreover, the adiabatic constraint is relaxed \cite{gorshkov2007photon2}, enabling shorter Gaussian pulses to be stored and retrieved perfectly, at the cost of requiring larger $g_{\Phi}$.

We further evaluate the optimized storage and retrieval $\eta$ and $\text{F}$ by separately considering the effects of pure dephasing $\gamma_{\phi}$ and loss $\Gamma_m$. The results for the resonant case, obtained under the same probe pulse condition as in Fig. \ref{fig:2}(a), are shown in Fig. \ref{fig:3}. Notably, the results for Raman storage exhibit the same behavior. The theoretical dependence of $\eta$ on $\gamma_{\phi}$ in Fig. \ref{fig:3}(a), given by $\eta=(\Gamma/2\gamma)^2$ (see Appendix \ref{AppendixC}), agrees well with the numerical results. In Fig. \ref{fig:3}(b), the decay of $\eta$ is governed by the coherence time of the metastable state, with fitting results confirming an exponential decay. Nevertheless, $\text{F}$ remains near unity even in the presence of either decoherence source.

\begin{figure}[t!]
    \centering
    \includegraphics[width=0.4\textwidth]{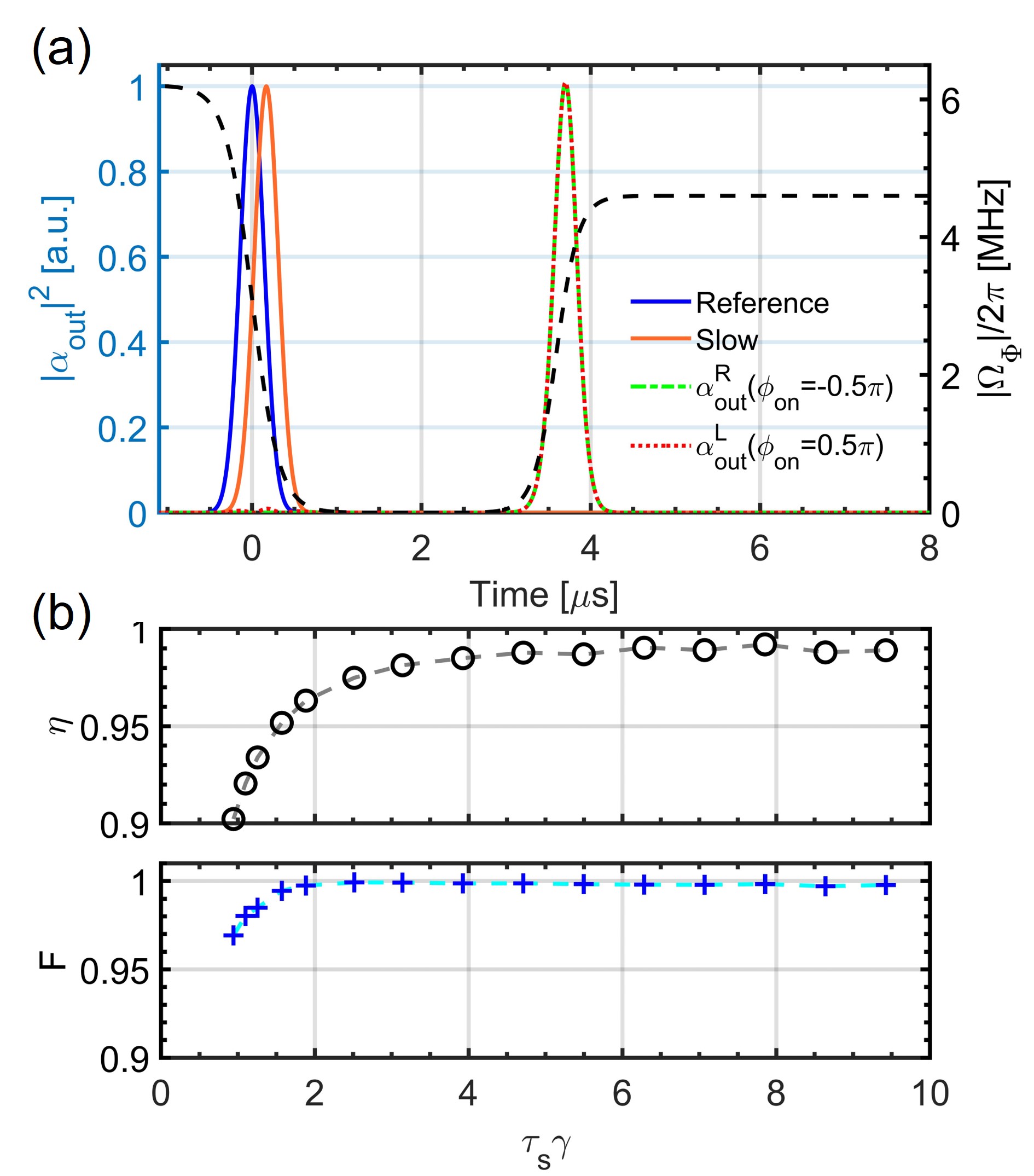}
    \caption{(a) Simple adiabatic control sequence. The black-dashed line shows the hyperbolic tangent control of the coupling field. The orange line denotes the slow probe pulse under constant $\Omega_{\Phi}$. The blue line represents the input probe pulse. The green and red-dotted lines depict the retrieved pulses propagating in opposite directions, achieved by tuning the turned-on phase difference $\phi_{\text{on}}$ of the two parametric modulations. (b) $\eta$ (black empty circles) and $\text{F}$ (blue cross symbols) as functions of $\tau_s$.}
    \label{fig: 4}
\end{figure}

The optimal control yields a function that explicitly depends on the shape of the input field to achieve perfect storage and retrieval. However, such a control function is rather complex and may require further characterization as well as compensation for potential distortions. To address this, we introduce a simpler EIT-based control protocol that can still achieve high storage and retrieval efficiency, while being more experimentally accessible. In this scheme, the coupling dynamics is governed by a hyperbolic tangent function, which adiabatically turns the coupling field off and on while the probe pulse is within the atomic medium, as shown in Fig. \ref{fig: 4}(a). By mapping $\eta$ and $\text{F}$ as functions of the adiabatic changing rate and the magnitude of $\Omega_{\Phi}$, the control parameters can be fine-tuned to achieve optimal performance (see Appendix \ref{AppendixE}). Fig. \ref{fig: 4}(a) illustrates the time evolution of the probe pulse with $\tau_s=200$ ns under the simple adiabatic control in the ideal system. Moreover, the retrieved pulse's propagation direction can be controlled by adjusting the phase difference $\phi_{on}$ of the turned-on parametric modulations. The left-propagating coherent field is calculated as $\hat{\alpha}^L_{out} =-i \sqrt{\Gamma/2} ( \sigma_1 + \sigma_2 e^{+i\pi/2} )$. The results show that the controllable $\phi_{\text{on}}$ induces an effect similar to the use of different propagation directions of the coupling light in atomic systems for controlling the direction of the retrieved pulse \cite{lin2009stationary}. $\eta$ and $\text{F}$ as functions of $\tau_s$ are plotted in Fig. \ref{fig: 4}(b), showing trends similar to Fig. \ref{fig:2}(b). A small oscillation of $\eta$ as a function of $\tau_s$ can be mitigated by slightly tuning the modulation turn-off time. In particular, for $\tau_s\gamma\approx1$, both quantities drop significantly. This can also be understood from the fact that the pulse spectral bandwidth overlaps with the Autler–Townes doublets of the excited state induced by the coupling field, leading to absorption at the resonance peaks and manifesting as quantum beating. In this non-adiabatic regime, an Autler-Townes-splitting-based quantum memory approach may help mitigate this issue and may also be applied to our chiral molecule system \cite{liao2014all,saglamyurek2018coherent}.

In conclusion, we propose a highly efficient $\Lambda$-type microwave memory based on a superconducting artificial chiral molecule embedded in an open transmission line. Using the optimal control function, our scheme achieves a storage and retrieval efficiency of nearly 100\% and near-unity fidelity, provided that the adiabatic condition is satisfied. The optimal control further allows the storage of arbitrary waveforms. Operating in Raman regimes, the storage bandwidth can reach values on the order of 100 MHz, making this system more practical for the quantum networks. The propagation direction of the retrieved pulse can also be precisely controlled. In addition, we demonstrate that the highly efficient EIT memory can be implemented with a simple adiabatic control protocol. Moreover, our method requires only a single memory qubit, avoiding the performance degradation associated with deviations in resonator array implementations \cite{bao2021demand,matanin2023toward,makihara2024parametrically}. By relying on a single chiral $\Lambda$-type system, our scheme contrasts with conventional atomic systems that depend on large optical depths. These results demonstrate a promising approach for achieving highly efficient microwave memory in superconducting circuits.

\begin{acknowledgments}
The work was supported by the National Science and Technology Council in Taiwan through Grants No. NSTC 113-2112-M-008-006, NSTC-110-2112-M-008-027-MY3, NSTC-111-2923-M-008-004-MY3 and NSTC-111-2639-M-007-001-ASP.
\end{acknowledgments}

\appendix
\section{system model}
\label{subsec:Method1}

\subsection{Parametric Modulation}
\label{appendixA1}

\begin{figure}[t!]
    \centering
    \includegraphics[width=0.49\textwidth]{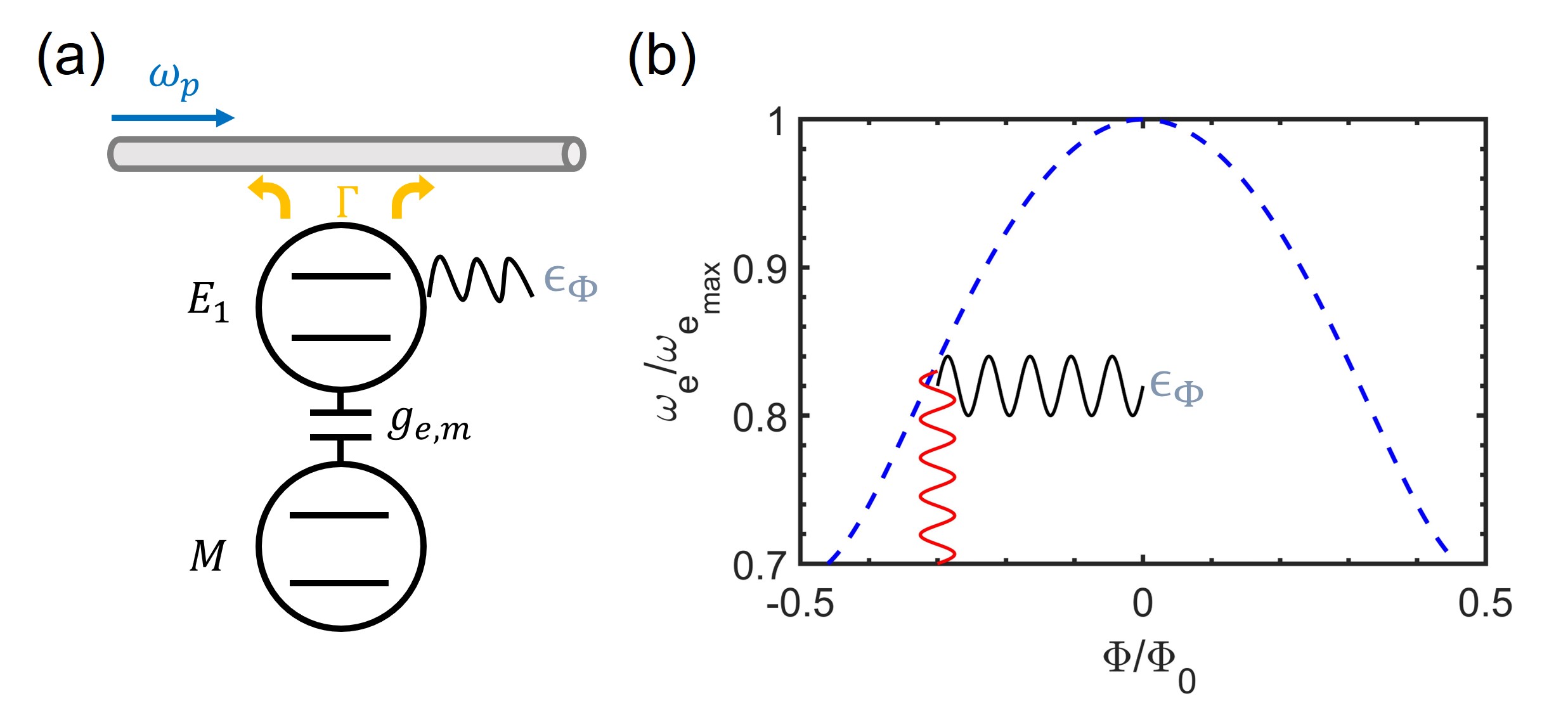}
    \caption{(a) The emitter is coupled with strength $g_{e,m}$ to a far-detuned memory qubit. Parametric modulation is applied to activate the otherwise forbidden transition between them.
(b) The emitter transition frequency as a function of flux. The emitter is periodically driven by a flux modulation (red), resulting in a modulation of its transition frequency (black).} 
    \label{fig:setup_appenedix}
\end{figure}

To activate the forbidden transition between the emitters and the memory qubit, one can apply parametric modulation to the emitter, the memory qubit, or even to the tunable coupler between them. Here, we consider the case where the modulation is applied to the emitter. To illustrate the effect of parametric modulation, we first focus on a simple scenario in which a single emitter is strongly coupled to a far-detuned memory qubit; see Fig. \ref{fig:setup_appenedix}(a). The probe field also interacts with the emitter. This scheme will later be extended to the chiral molecule.
Fig. \ref{fig:setup_appenedix}(b) illustrates the parametric modulation, where the emitter transition frequency is biased in the linear regime and periodically modulated at frequency $\omega_{\Phi}$, expressed as
\begin{equation}
    \label{eq:parametric drive}
    \omega_{e}(t) \approx \omega_{e} + \frac{\epsilon_{\Phi}}{2}\sin{(\omega_{\Phi}t)},
\end{equation} 
where $\epsilon_{\Phi}$ denotes the modulation strength of the transition frequency. The Hamiltonian of the system can then be written as 
 \begin{equation}
 \begin{aligned}
    \label{eq:Driving Hamiltonian}
    H=&[\omega_{e}(t)-\omega_{p}]\sigma_{1}^\dagger\sigma_{1} + (\omega_{m}-\omega_{p})\sigma_m^\dagger \sigma_m \\&+ g_{e,m}(\sigma_m^\dagger\sigma_1 + \sigma_1^\dagger \sigma_m)+\frac{\Omega_{p}}{2}(\sigma_1^\dagger+\sigma_1),
\end{aligned}
\end{equation}
where the rotating frame at $\omega_{p}$ has been applied and the fast-rotating terms have been neglected. To capture the effect of parametric modulation, the unitary transformation $U^a_{\text{rot}}$ in the rotating frame of the instantaneous oscillation frequency is further applied, defined as 
\begin{equation}
\begin{aligned}
    U^a_{rot} = \exp\bigg[&i(\Delta_p t - \frac{\epsilon_{\Phi}}{2\omega_{\Phi}}\cos{\omega_{\Phi}t})\sigma_1^\dagger\sigma_1\\
    &+ i(\omega_m-\omega_p)t \sigma_m^\dagger \sigma_m \bigg],
\end{aligned}
\end{equation}
where $\Delta_p=\omega_e-\omega_p$ is the one-photon detuning. The resulting Hamiltonian (denoted as $H$ for simplicity) is given by 
\begin{equation}
    \label{eq: 1Hamiltonian}
    \begin{aligned}
    H =&g_{e,m}\sum_{n=-\infty}^{\infty}J_{n}(\frac{\epsilon_{\Phi}}{2\omega_{\Phi}})[i^n e^{i(n\omega_{\Phi}-\Delta_{e,m})t}\sigma_m^\dagger\sigma_1 + \textrm{H.c.}]\\&+\frac{\Omega_p}{2}[e^{-i\Delta_pt}\sigma_1+\textrm{H.c.}],
    \end{aligned}
    \end{equation}
where $\Delta_{e,m}=\omega_e-\omega_m$ and $J_n$ denotes the Bessel function of the first kind. Here, the condition $\epsilon_{\Phi}\ll 2\omega_{\Phi}$ is imposed to ensure that $\Omega_p$ is not affected by the parametric modulation \cite{chu2023three}. For resonant modulation, the forbidden transition between the emitter and the memory qubit is activated with an effective coupling strength
\begin{equation}
    \label{eq:Parametric rabi}
g_{\Phi}=2g_{e,m}J_{1}(\frac{\epsilon_{\Phi}}{2\omega_{\Phi}}).
\end{equation}
Care must be taken with the nonlinearity of $J_{1}$ under strong modulation, and the maximum $g_{\Phi}$ is approximately equal to $1.16g_{e,m}$.

An additional unitary transformation is employed to eliminate the time-dependent terms in Eq. \ref{eq: 1Hamiltonian}, expressed as 

\begin{equation}
    \label{eq:Unitary operator2}
    U^b_{rot} = \exp(-i\Delta_pt\sigma_1^\dagger\sigma_1-i\delta t\sigma_m^\dagger \sigma_m),
\end{equation}
where $\delta=\omega_m-\omega_p+\omega_{\Phi}=\Delta_p-\Delta_{\Phi}$ is the two photon detuning.
After neglecting the fast-rotating terms, the time-independent Hamiltonian can be written as
\begin{equation}
\begin{aligned}
\label{eq:Final Hamiltonian}
H=&\Delta_p\sigma_1^\dagger\sigma_1+\delta \sigma_m^\dagger \sigma_m\\&+ g_{\Phi}(\sigma_m^\dagger\sigma_1+\sigma_m\sigma_1^\dagger)+\frac{\Omega_p}{2}(\sigma_1+\sigma^\dagger_1),
\end{aligned}
\end{equation}
where the phase factor has been absorbed into $\sigma_m$.

\subsection{Derivation of Hamiltonian}
\label{AppendixA2}


In this section, we start from the interaction between the molecule and the bath and derive the Hamiltonian used in the main text. We then examine the collective effect between the emitters mediated by the transmission line, which involves the free-evolution approximation and the slowly-varying-decay-rate approximation. In other words, the Markovian approximation is applied. The original derivation of the collective effect can be found in Ref.~\cite{gheeraert2020programmable}. These approximations ensure that the atoms experience the same probe-field amplitude but with a propagating phase difference, leading to the main Hamiltonian and the input–output relations used in the main text.

Next, we introduce a second emitter into the system that is also coupled to the open transmission line, and allow both emitters to interact with the memory qubit. A tunable coupler is introduced to cancel the collective coupling Hamiltonian $H_J$ mediated through the waveguide, which will be derived later. Note that the total dispersive coupling can also be engineered to cancel the collective coupling.

The molecular Hamiltonian of the system is written as
\begin{equation}
\begin{aligned}
    \label{eq:molecule}
H_m &=\sum_{i=1,2} \left( \omega_e(t)\sigma_{i}^\dagger \sigma_{i} + g_{e,m}(\sigma_m^\dagger\sigma_i + \sigma_i^\dagger \sigma_m) \right)+\omega_{m}\sigma_m^\dagger \sigma_m\\&+H_c,
\end{aligned}
\end{equation}
where $H_c=-\frac{\Gamma_1}{2}(\sigma_1^\dagger\sigma_2+\sigma_2^\dagger\sigma_1)$ is the coupling through the tunable coupler.

The Hamiltonian of the bath in the open transmission line is expressed as 
\begin{equation}
    \label{eq:bath}
H_b=\int_{-\infty}^{\infty}dkk b_{R}^\dagger(k) b_{R}(k) + \int_{-\infty}^{\infty}dkk b_{L}^\dagger(k) b_{L}(k),
\end{equation}
where $b_{R(L)}$ denotes the lowering operator of the right (left) propagating mode and $k$ is the wave vector. The operators satisfy the following commutation relation $[b_{R}(k),b_{R}^\dagger(k')]=[b_{L}(k),b_{L}^\dagger(k')]=\delta(k-k')$.
The interaction Hamiltonian between the bath and the molecule is given by
\begin{equation}
\begin{aligned}
    \label{eq:int_m_b}
H_{int}=&\sum_{i=1,2}\int_{-\infty}^{\infty}\sqrt{\frac{\Gamma_i}{2}}\frac{dk}{\sqrt{2\pi}}\biggl(\sigma_i^\dagger b_R(k)e^{ikr_i}  + \sigma_i b_R(k)^\dagger e^{-ikr_i} \\&+ \sigma_i^\dagger b_L(k)e^{-ikr_i} + \sigma_i b_L(k)^\dagger e^{ikr_i} \biggr),
  \end{aligned}
\end{equation}
where $\Gamma_{1(2)}$ and $r_{1(2)}$ denote the relaxation rate and the position of the emitter 1(2), respectively. Here, the field propagation velocity is set to unity.

We can express the Heisenberg equation of motion for an arbitrary operator $A$ of the molecule–bath system as
\begin{equation}
\begin{aligned}
\label{eq:operatorA}
\frac{dA}{dt}
&= i[H_m+H_b,A] \\
&\; + i\sum_{i=1,2}\!\int_{-\infty}^{\infty}\!\sqrt{\frac{\Gamma_i}{2}}\frac{dk}{\sqrt{2\pi}}
\biggl( [\sigma_i^\dagger b_R,A]e^{ikr_i} + e^{-ikr_i}[b_R^\dagger\sigma_i,A] \\
&\;\; + [\sigma_i^\dagger b_L,A]e^{-ikr_i} + e^{ikr_i}[b_L^\dagger\sigma_i,A] \biggr).
\end{aligned}
\end{equation}
The corresponding solution of $b_R$ in the Heisenberg picture is given by
\begin{equation}
    \begin{aligned}
    \label{eq:br}
b_R(k)&=b_R(k)e^{-ik(t-t_i)}\\
&-\frac{i}{\sqrt{2\pi}}\sum_{i=1,2}\int_{t_i}^{t}dt'\sqrt{\frac{\Gamma_i(t')}{2}}\sigma_i(t')e^{-ik(r_i-t'+t)},
    \end{aligned}
\end{equation}
where $t_i$ denotes the initial time.

The field operator in $k$-space can be converted to the position-space representation by integrating over all wave vectors. The right- and left-propagating field operators $\hat\alpha^{R(L)}_r$ at position $r$ are defined as 

\begin{subequations}
\label{eq:field at r}
\begin{align}
    \hat\alpha^R_r & =\int_{-\infty}^{\infty}\frac{dk}{\sqrt{2\pi}}b_R(k)e^{ikr},
    \\
    \hat\alpha^L_r & =\int_{-\infty}^{\infty}\frac{dk}{\sqrt{2\pi}}b_L(k)e^{-ikr}.
\end{align}
\end{subequations}
The corresponding time-evolved forms are obtained as
\begin{widetext}
\begin{subequations}
\label{eq:field at r2}
\begin{align}
\hat\alpha^R_r = \hat\alpha^R_{r+t_i-t}(t_i)-i\sum_{i=1,2}\sqrt{\frac{\Gamma_i(t-r+r_i)}{2}}\sigma_i(t-r+r_i)\Theta(r-r_i)\Theta(r_i+t-r),
\\
\hat\alpha^L_r = \hat\alpha^L_{r+t-t_i}(t_i)-i\sum_{i=1,2}\sqrt{\frac{\Gamma_i(t+r-r_i)}{2}}\sigma_i(t+r-r_i)\Theta(r-r_i+t)\Theta(r_i-r),
\end{align}
\end{subequations}
\end{widetext}
where $\Theta$ is the step function, which enforces the time ordering of events. The position-space field operators satisfy the commutation relations  $[\hat\alpha^R_r,\hat\alpha^{R\dagger}_{r'}]=[\hat\alpha^L_r,\hat\alpha^{L\dagger}_{r'}]=\delta(r-r')$.

We now introduce the free-evolution approximation $\sigma_i(t-d)\approx\sigma_ie^{i\omega_ed}$, where $d$ is the propagation distance (i.e., the field delay) between the two emitters, and the slowly-varying-decay-rate approximation $\gamma_i(t-d)\approx\gamma_i(t)$. These approximations hold when the characteristic evolution time of the system $1/\Gamma$ is much longer than the propagation delay $\tau_p\equiv d$, namely when $\Gamma \tau_p \ll 1$. As $\Gamma \tau_p$ approaches unity, the approximation breaks down and non-Markovian effects associated with retardation begin to appear. In this regime, the delayed field emitted by the first atom can influence the dynamics of the second atom~\cite{guo2017giant,sinha2020non}. In addition, we have assumed that the coupling strength to the waveguide satisfies $g_w \ll \omega_e$, such that the rotating wave approximation and Born-Markovian approximations remain valid~\cite{Breuer2016,Combes2017,Maffei2024}.

We further consider a right-propagating classical drive field $\hat\alpha^R_{in}(r,t)$. The field equation can then be written as
\begin{subequations}
\label{eq:field equation}
\begin{align}
\hat\alpha^R_{r+t_i-t}(t_i)=\hat\alpha^R_{in}(r,t)+\xi^R_{vac}
    \\
\hat\alpha^L_{r-t_i+t}(t_i)=\xi^L_{vac},
\end{align}
\end{subequations}
where $\xi^{R(L)}_{vac}$ is the right (left) vacuum mode.
Because there is no feedback from the molecule to the bath under these approximations (also see the discussion in main text), the classical field amplitudes at different atomic positions are related as $\hat\alpha^R_{in}(r_2,t)=\hat\alpha^R_{in}(r_1,t-d)\approx\hat\alpha^R_{in}e^{i\omega_dd}$, where $\omega_d$ is the drive frequency. We substitute Eq. \ref{eq:field at r}, \ref{eq:field at r2}, and \ref{eq:field equation} into the Heisenberg equation of the atomic operator $O$ in the molecule subspace in the position form (where the field operator is no longer in the commutator in Eq. \ref{eq:operatorA}), we obtain

\begin{equation}
    \begin{aligned}
    \label{eq:operator O}
    \frac{dO}{dt} & = i[H_m,O] + i\sum_{i=1,2}([\sigma_i^\dagger,O]\alpha_{in}^R(r_i,t)+\alpha_{in}^{R*}(r_i,t)[\sigma_i,O])
    \\&+\sum_{i=1,2}\frac{\Gamma_i}{2}([\sigma_i^\dagger,O]\sigma_i-\sigma_i^\dagger[\sigma_i,O])
    \\&+\sum_{i=1,2}\frac{\sqrt{\Gamma_1\Gamma_2}}{2}([\sigma_i^\dagger,O]\sigma_{\bar{i}}e^{i\omega_ed}-\sigma_{\bar{i}}[\sigma_i,O]e^{-i\omega_ed}).
    \end{aligned}
\end{equation}
The second term in Eq. \ref{eq:operator O} corresponds to the driving Hamiltonian acting on the molecule,
\begin{equation}
    \begin{aligned}
    \label{eq:drive}
H_d &= \sum_{i=1,2}\sqrt{\frac{\Gamma}{2}}(\alpha_{in}^R(r_i,t)\sigma_i^\dagger+\alpha_{in}^{R*}(r_i,t)\sigma_i)
\\& =\frac{\Omega_p}{2}(\sigma_1^\dagger + \sigma_1 +e^{ikd}\sigma_2^\dagger +e^{-ikd}\sigma_2),
    \end{aligned}
\end{equation}
where we assume $\Gamma_1=\Gamma_2=\Gamma$.

Rewriting the above equations in the Schrödinger picture, the time evolution of the reduced density matrix of the molecule is obtained as
\begin{equation}
    \begin{aligned}
    \label{eq:final}
\frac{d\rho}{dt}=-i[H_m+H_d+H_J,\rho]+\sum_{i,j=1,2}\Gamma_{ij}\mathcal{D'}(\sigma_i,\sigma_j)\rho,
    \end{aligned}
\end{equation}
where the collective coupling through the waveguide is given by $H_J=J(\sigma_1^\dagger\sigma_2+\sigma_2^\dagger\sigma_1)$, with $J=\frac{\Gamma}{2}\sin{\omega_ed}$. The superoperator describing the correlated decay is defined as $\mathcal{D'}(\sigma_i,\sigma_j)=\sigma_j\rho\sigma_i^\dagger- \tfrac{1}{2}\{\sigma_i^\dagger\sigma_j, \rho\}_{+}$. We set $\Gamma_{11}=\Gamma_{22}=\Gamma$, while the collective decay rates are given by $\Gamma_{12}=\Gamma_{21}^*=\Gamma\cos{\omega_ed}$. In our setup, since $\omega_ed=\pi/2$, the system exhibits no collective decay but achieves maximum collective coupling, which is canceled by the tunable coupler $H_c$.

Furthermore, under these approximations, the input–output relation takes a simple form,
\begin{equation}
\begin{aligned}
    \label{eq:rfiled}
\hat{\alpha}^R_{out}= \hat{\alpha}^R_{in} - i\sqrt{\frac{\Gamma}{2}} ( \sigma_1 + \sigma_2 e^{-i\frac{\pi}{2}} ),
\end{aligned}
\end{equation}
where we define $\hat\alpha^R_r=\hat{\alpha}^R_{out}$ and $\hat\alpha^R_{r+t-t_i}=\hat\alpha^R_{in}$.

\subsection{Chiral Molecule}
\label{AppendixA3}

 Assuming that the total coupling strength between the emitters vanishes, and applying the modulations with different phases to the two emitters, the Hamiltonian can be written as (see main text Eq. 1),
\begin{equation}
    \begin{aligned}
    \label{eq:H1}
    H = &\sum_{i=1,2} \Delta_p \sigma_{i}^\dagger \sigma_{i} \\ &+ g_{\Phi}^*(e^{-i\frac{\pi}{2}} \sigma_{1}^\dagger \sigma_{m} + \sigma_{2}^\dagger \sigma_{m}) + \frac{\Omega_p}{2} (\sigma_{1}^\dagger + e^{i\frac{\pi}{2}} \sigma_{2}^\dagger) + \text{H.c.},
    \end{aligned}
\end{equation}
where we consider two-photon resonant $\delta=0$, and $kd=\pi/2$ and $\phi=-\pi/2$. To illustrate the chiral property, we introduce the right- and left-chiral operators, expressed as

\begin{subequations}
\begin{align}
    \label{eq:chiral operator}
    \sigma_R & = \frac{1}{\sqrt{2}}(\sigma_1-i\sigma_2),
    \\
    \sigma_L & = \frac{1}{\sqrt{2}}(\sigma_1+i\sigma_2).
\end{align}
\end{subequations}
Rewriting the Hamiltonian Eq. \ref{eq:H1} with the chiral operators and absorbing the phase factor into $\sigma_m$, we obtain
\begin{equation}
    \begin{aligned}
    \label{eq:H2}
    H = &\Delta_p( \sigma_{R}^\dagger \sigma_{R}+\sigma_{L}^\dagger \sigma_{L}) \\ &
    +\sqrt{2}(g_{\Phi}^*\sigma_{m}^\dagger\sigma_{R}+g_{\Phi}\sigma_{R}^\dagger\sigma_{m})+\frac{\Omega_p}{\sqrt{2}}(\sigma_{R}^\dagger+\sigma_{R}). 
    \end{aligned}
\end{equation}
The two modulations can be regarded as a coupling field for the chiral molecule, with the coupling Rabi frequency defined as $\Omega_{\Phi}=2\sqrt{2}|g_{\Phi}|$. From Eq. \ref{eq:H2}, one finds that the right-going resonant probe field couples exclusively to the right-chiral operator when $\phi=-\pi/2$, which leads to the chiral effect.

\subsection{Effective $\Lambda$-Type System \& Dark State}
\label{appendixA4}

In the weak-probe regime, we use the lowest three dressed states, which form a $\Lambda$-type structure describing the chiral molecule (see Fig. 1 in the main text): the ground state $\ket{G}=\ket{gg}\otimes\ket{0}$, the excited state $\ket{E}=(\ket{eg} + e^{i\pi/2} \ket{ge})/\sqrt{2} \otimes \ket{0}$, and the metastable state $\ket{M}=\ket{gg}\otimes\ket{1}$, where $\ket{0}$ and $\ket{1}$ denote the ground and excited states of the memory qubit, respectively. With the probe and modulation fields satisfying the two-photon resonance condition, we observe the dark state with zero energy, written as
\begin{equation}
    \label{eq:D}
    \ket{D}=\frac{1}{\sqrt{\frac{|\Omega_{\Phi}|^2}{2}+|\Omega_p|^2}}(\frac{\Omega_{\Phi}}{\sqrt{2}}\ket{G}-\Omega_p\ket{M}).   
\end{equation}
This dark state corresponds to that of a $\Lambda$-type optical memory in AMO systems, where the probe field can be mapped onto the state $\ket{M}$ by adiabatically turning off the modulation $\Omega_{\Phi}$.

We can also formulate the Heisenberg–Langevin equations together with the input–output relations for the system:
\begin{subequations}
\label{eq:HL}
\begin{align}
    \dot{\sigma_R}&=-(i\Delta_p+\gamma)\sigma_R-i\frac{\Omega_{\Phi}}{2}\sigma_M-i\sqrt{\Gamma}\hat{\alpha}^R_{in},
    \\
    \dot{\sigma_M} & = -i\frac{\Omega_{\Phi}^*}{2}\sigma_R,
    \\
    \hat{\alpha}^R_{out}& = \hat{\alpha}^R_{in} - i\sqrt{\Gamma} \sigma_R,
\end{align}
\end{subequations}
where we assume a perfect memory qubit, and $\gamma=\gamma_{\phi}+\Gamma/2$ denotes the total decoherence rate of the emitters. Eq. \ref{eq:HL} generally describes the equation of motion for any chiral $\Lambda$-type system within the waveguide quantum electrodynamics architecture.

\section{Steady state Response \& slow light}
\label{AppendixB}

\begin{figure}[t!]
    \centering
    \includegraphics[width=0.49\textwidth]{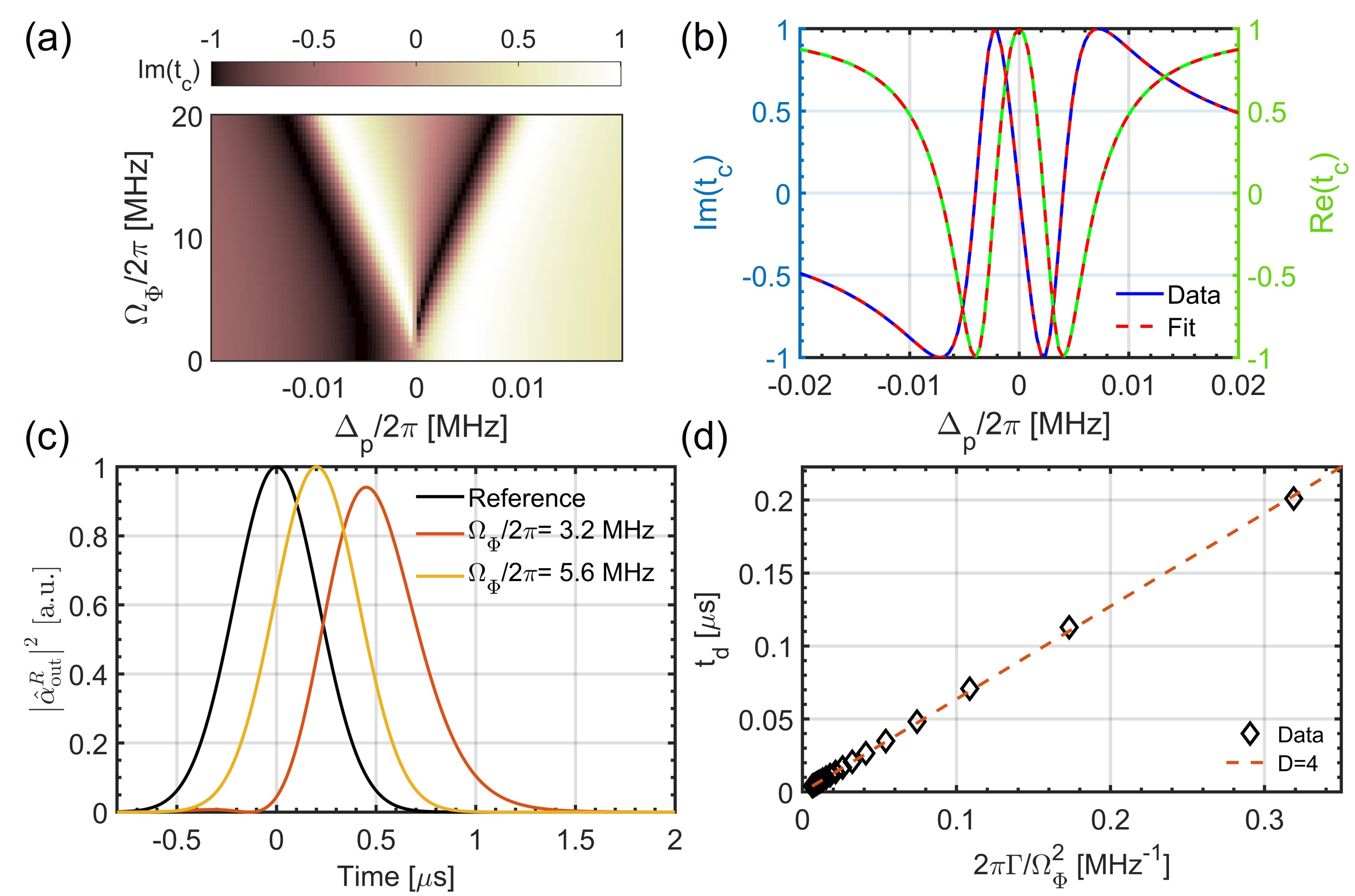}
    \caption{(a) The imaginary part of the probe transmission coefficient Im$\left( t_c\right) $ as a function of  $\Delta_p$ and $\Omega_{\Phi}$. (b) Blue-solid (Green-solid) line depicts $\Delta_p$-dependent imaginary (real) part of $t_c$ for $\Omega_{\Phi}/2\pi = 8$ MHz. Dashed lines represent the theoretical curve. 
    (c) Slow probe light signal $|\hat{\alpha}^R_{out}|^2$ for $\Omega_{\Phi}/2\pi=3.2$ MHz (yellow-solid line) and 5.6 MHz (red-solid line). The black-solid line is the input probe pulse. (d) Group delay time $t_d$ as a function of $\Omega_{\Phi}$. The black open diamonds represent data extracted from (c), and the red dashed line illustrates the EIT delay-time formula $4\Gamma/\Omega_{\Phi}^2$.}
    \label{fig:EIT}
\end{figure}

In this section, we analyze the EIT mechanism and the slow-light effect under the assumption of an ideal system with no dephasing or loss. The time evolution of the system can be obtained by solving the Lindblad master equation in combination with the input–output theory, as given in the main text, which allows us to characterize the optical response of the system. The steady-state transmission coefficient is defined as $t_c=\hat{\alpha}^R_{\text{out}}/\hat{\alpha}^R_{\text{in}}$. The real and imaginary parts of $t_c$ are plotted as a function of the probe detuning $\Delta_p$ with resonant coupling light in Fig. \ref{fig:EIT}(b). The data can be well described by the following general analytical expression for the transmission coefficient \cite{chu2023three}:
\begin{equation}
    \label{eq:Driven Lambda type}
    t^{a}_{c} = 1 + 2i\frac{\frac{\Gamma}{2}({\Delta_p-i\frac{\Gamma_m}{2})}}{(\Delta_p-i\frac{\Gamma_m}{2})(\Delta_p-i\gamma)-(\frac{\Omega_{\Phi}^2}{4})},
\end{equation}
By fitting the data in Fig. \ref{fig:EIT}(a) and (b) to Eq. \ref{eq:Driven Lambda type}, we find that the condition $\Omega_{\Phi} = 2\sqrt{2}|g_{\Phi}|$ holds, where Fig. \ref{fig:EIT}(a) illustrates how $\text{Im}(t_c)$ varies with $\Delta_p$ and $\Omega_{\Phi}$. This tunable dispersion indicates that the group delay $t_d$ of the probe can be controlled by adjusting $\Omega_{\Phi}$.

To observe the EIT slow-light effect, we set $\Delta_p=\delta=0$. From the dark state in Eq. \ref{eq:D}, one finds the relation $\sigma_m=-\sqrt{2}\Omega_p/\Omega_{\Phi}$. Applying this condition to Eq. \ref{eq:HL}(b) together with $\Omega_p=\sqrt{2\Gamma}\hat{\alpha}^R_{in}$, we derive
\begin{equation}
    \label{eq:sl1}
    \sigma_R=-i\frac{4\sqrt{\Gamma}}{|\Omega_{\Phi}|^2}\dot{\hat{\alpha}}^R_{in}.
\end{equation}
Substituting Eq.\ref{eq:sl1} into Eq.\ref{eq:HL}(c), we obtain the slow-light effect, given as
\begin{equation}
    \begin{aligned}
    \label{eq:sl2}
     \hat{\alpha}^R_{out}& = \hat{\alpha}^R_{in}-\frac{4\Gamma}{|\Omega_{\Phi}|^2}\dot{\hat{\alpha}}^R_{in}
     \\& \approx\hat{\alpha}^R_{in}(t-t_d),
    \end{aligned}
\end{equation}
where $t_d$ denotes the group delay time, which depends on $\Omega_{\Phi}$. We further measure the slow-light effect by sending a Gaussian probe pulse, $\Omega_p\exp(-t^2/2{\tau_s}^2)$, with a duration of $\tau_s=300$ ns into the molecule under continuous coupling light. The result is shown in Fig. \ref{fig:EIT}(c). As $\Omega_{\Phi}$ decreases, $t_d$ increases due to the steepening of the dispersion. To determine the effective optical depth $D$ of the chiral system, we analyze $t_d$ as a function of $\Omega_{\Phi}$ in Fig. \ref{fig:EIT}(d) and compare it with the expression $t_d = D\Gamma/\Omega_{\Phi}^2$, commonly used in AMO systems. Both analytical and numerical results indicate that the optical depth increases to $D=4$, in contrast to previous work \cite{chu2023three,chu2025slow}.

\section{Efficiency independent of optical depth}
\label{AppendixC}

In this section, we show that the storage and retrieval efficiency $\eta$ is independent of the optical depth in the chiral $\Lambda$-type system within waveguide quantum electrodynamics, in sharp contrast to AMO systems. Moreover, $\eta$ is also independent of $\Delta_p$ and $\Omega_{\Phi}$. The calculation follows a procedure similar to Ref.\cite{gorshkov2007photon2}. We assume perfect storage, where $\sigma_m$ is initially prepared. During the retrieval process, the last term in Eq. \ref{eq:HL}(a) is set to zero since there is no input field. From Eqs. \ref{eq:HL}(a) and (b), one finds the time derivative of the total coherence,
\begin{equation}
    \begin{aligned}
    \label{eq:ef1}
    \frac{d}{dt}(|\sigma_R|^2+|\sigma_M|^2)=-2\gamma|\sigma_R|^2.
    \end{aligned}
\end{equation}
Integrating Eq. \ref{eq:ef1} from the retrieval time $t_r$ to infinity and assuming $\sigma_R(\infty)=\sigma_M(\infty)=0$, we obtain
\begin{equation}
    \begin{aligned}
    \label{eq:ef2}
|\sigma_M(t_r)|^2=2\gamma\int_{t_{r}}^{\infty}|\sigma_R|^2dt.
    \end{aligned}
\end{equation}
By also integrating Eq. \ref{eq:HL}(c) (with $\hat{\alpha}^R_{in}=0$), the retrieval efficiency $\eta_r$ can be expressed as
\begin{equation}
    \begin{aligned}
    \label{eq:ef3}
\eta_r&=\frac{\int_{t_{r}}^{\infty} |\hat{\alpha}^R_{out}|^2 dt}{|\sigma_M|^2}
\\&=\frac{\Gamma}{2\gamma}.
    \end{aligned}
\end{equation}
Thus, $\eta_r$ is limited only by the pure dephasing rate. The only requirement is that $\Omega_{\Phi}$ is sufficiently strong to ensure that no residual coherence remains in the system. Owing to time-reversal symmetry \cite{gorshkov2007photon2}, the maximum storage efficiency equals $\eta_r$, and the total storage and retrieval efficiency is therefore $\eta=(\Gamma/2\gamma)^2$.

\section{Derivation of Optimal Control}
\label{AppendixD}

\begin{figure}[t!]
    \centering   \includegraphics[width=0.35\textwidth]{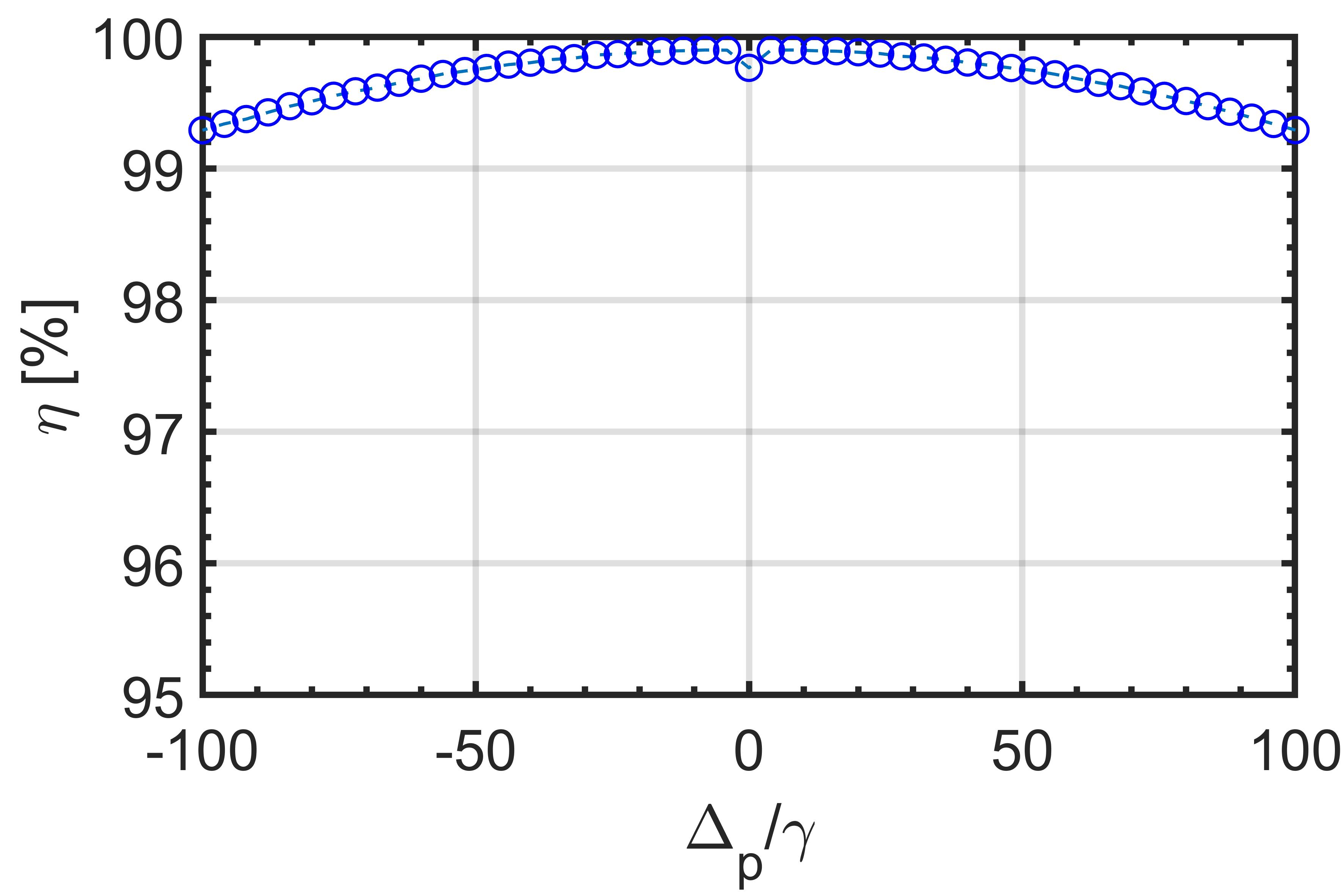}
    \caption{Optimal $\eta$ of an ideal system as a function of $\Delta_p$.} 
    \label{fig:Raman}
\end{figure}

We consider the retrieval process under two conditions to derive the optimal control function. The first condition is the temporal retrieval-matching condition, $\hat{\alpha}^R_{out}(t)=\sqrt{\eta}\hat{\alpha}^R_{in}(t-t_r)$, so that the input–output relation in Eq. \ref{eq:HL}(c) imposes the constraint
\begin{equation}
\begin{aligned}
    \label{eq:opt0}
\sigma_R=i\frac{1}{\sqrt{\Gamma}}\hat{\alpha}^R_{ret},
\end{aligned}
\end{equation}
where $\hat{\alpha}^R_{ret}=\sqrt{\eta}\hat{\alpha}^R_{in}$. Substituting Eq. \ref{eq:opt0} into Eq. \ref{eq:HL}(a) and assuming the adiabatic condition $\dot{\sigma_R}\approx0$, we obtain
\begin{equation}
\begin{aligned}
    \label{eq:opt1}
\dot{\sigma_M}=-\frac{2(i\Delta_p+\gamma)}{\sqrt{\Gamma}}\frac{d}{dt}(\frac{\hat{\alpha}^R_{ret}}{\Omega_{\Phi}}).
\end{aligned}
\end{equation}
Substituting Eq. \ref{eq:opt0} and \ref{eq:opt1} into Eq. \ref{eq:HL}(b), one can solve for the optimal retrieval function (ignoring the global phase $\ln\eta$), written as 
\begin{equation}
\begin{aligned}
    \label{eq:opt}
\Omega_{\Phi}(t)=-\sqrt{2}\frac{i\Delta_p+\gamma}{\sqrt{\gamma}}\frac{\hat{\alpha}^R_{in}(t-t_r)}{\sqrt{S_{r}}} e^{i\frac{\Delta_p}{2\gamma}\ln{S_{r}}},
\end{aligned}
\end{equation}
where $S_{r}=\int_{t}^{\infty} |\hat{\alpha}^R_{in}(\tau-t_r)|^2 d\tau$. The time reversal of this optimal retrieval control corresponds to the optimal storage control.

For the detuned Raman memory, one requires the phase modulation to the optimal control to compensate for the AC Stark shift effect due to the large detuning. We further study the storage bandwidth $\Delta_P$ with a $\tau_s=400$ ns Gaussian pulse. Since Eq. \ref{eq:HL} describes the dynamics of the ideal chiral $\Lambda$-type system, the numerical simulation is instead based on the Hamiltonian and the input–output theory in the main text to account for the effect of phase matching. The result is shown in Fig. \ref{fig:Raman}. With $\Delta_p=100\gamma$, the efficiency $\eta$ remains above 99 \%. The slight nonlinearity at zero detuning arises because the adiabatic condition becomes less restrictive when the detuning deviates from zero.

For experiments, one should note that the maximum coupling is $g_{\Phi}\approx1.16g_{e,m}$, and particular care must be taken regarding nonlinear effects under strong modulation.

\begin{figure}[!t]
    \centering   \includegraphics[width=0.49\textwidth]{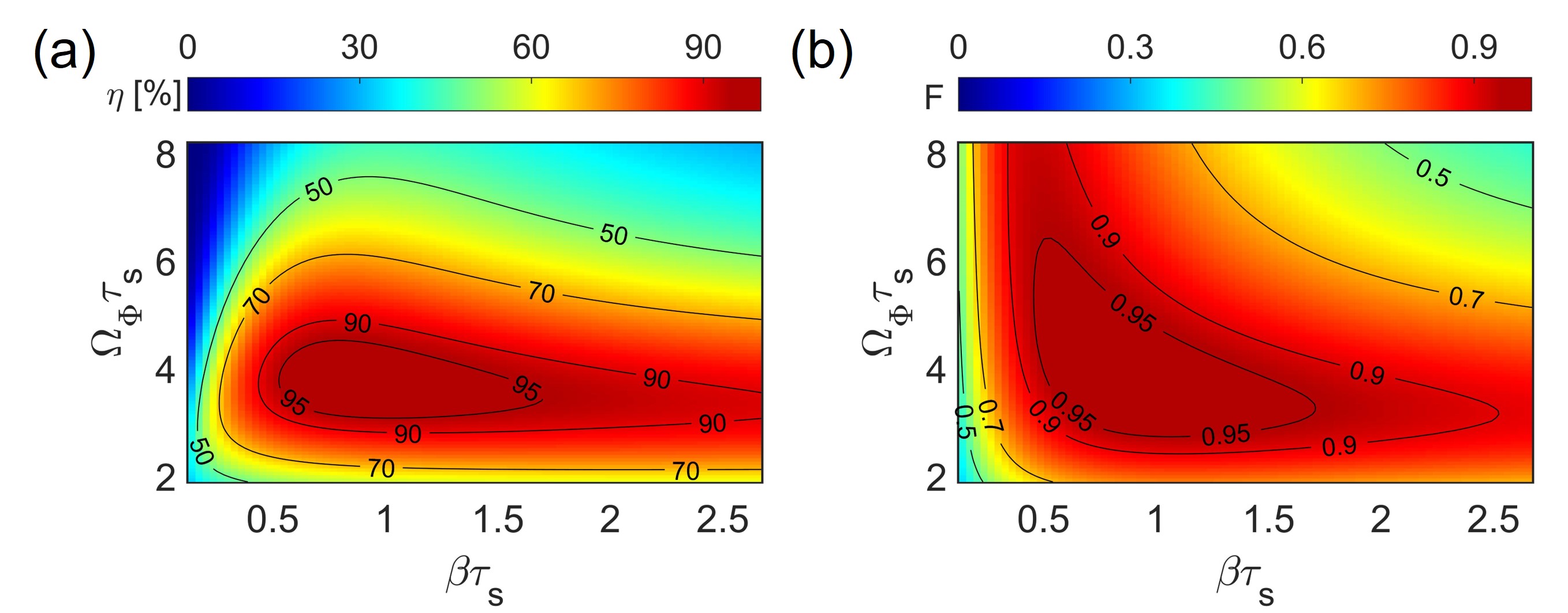}
    \caption{(a) Storage efficiency $\eta$ and (b) fidelity $\text{F}$ as functions of $\Omega_{\Phi}$ and switching slope $\beta$.} 
    \label{fig:storage_opt}
\end{figure}

\subsection{Adiabatic Condition}
\label{AppendixD1}
In the previous discussion, we introduced the adiabatic condition $\dot{\sigma_R}\approx0$ which implies $|\dot{\sigma_R}|\ll|(i\Delta_p+\gamma)\sigma_R|$ \cite{gorshkov2007photon2}. This condition imposes additional constraints on the power and bandwidth of the coupling and probe fields, following the analysis in Ref.~\cite{gorshkov2007photon2}. For the retrieval process, the condition yields

\begin{subequations}
\label{eq:ac}
\begin{align}
    |\frac{\Omega_{\Phi}}{2}|\ll|(i\Delta_p+\gamma)|,
    \\
    |\frac{\dot{\Omega_{\Phi}}}{\Omega_{\Phi}}|\ll|(i\Delta_p+\gamma)|.
\end{align}
\end{subequations}
The first inequality corresponds to the power constraint, while the second corresponds to the bandwidth constraint. In the resonant case, excessively strong coupling leads to Autler–Townes splitting rather than an interference-type dark state.

For the storage process, the requirement is
\begin{subequations}
\label{eq:ac2}
\begin{align} 
    |\frac{\dot{\hat{\alpha}}^R_{in}}{\hat{\alpha}^R_{in}}|\ll|(i\Delta_p+\gamma)|.
\end{align}
\end{subequations}
For a smooth Gaussian pulse, this condition reduces to the probe-pulse bandwidth constraint $1/\tau_s \ll |i\Delta_p+\gamma|$.

\section{Simple adiabatic Control}
\label{AppendixE}
We employ a simple adiabatic control function to study the optimization of EIT-based light storage. The control function is defined as
\begin{equation}
    \label{eq:control}
\frac{\Omega_{\Phi}}{2}([1-\tanh{\beta(t-t_{off})}]+[1+\tanh{\beta(t-t_{on}})]),
\end{equation}
where $\beta$, $t_{off}$, and $t_{on}$ represent the slope, turn-off time, and turn-on time, respectively. Figure~\ref{fig:storage_opt} shows the efficiency $\eta$ and the fidelity $F$ for a probe pulse of duration $\tau_s=100$ ns with $t_{\text{off}}=80$ ns, plotted as functions of $\Omega_{\Phi}$ and $\beta$. The optimized values of $\eta$ and $\text{F}$ indicate how to fine-tune the turn-off parameters $\Omega_{\Phi}$ and $\beta$, as well as the turn-on parameters, to achieve optimal storage and retrieval performance. In the main text, we fix $t_{\mathrm{off}} = 0$ to reduce the number of free parameters. In practice, different pulse durations would require different $t_{\mathrm{off}}$, but the efficiency is only slightly affected.


%

\end{document}